# Direct writing of heterostructures in single atomically precise graphene nanoribbons

Chuanxu Ma,[1,#] Zhongcan Xiao,[2,#] Jingsong Huang,[1,3] Liangbo Liang,[1] Wenchang Lu,[2,3] Kunlun Hong,[1] Bobby G. Sumpter,[1,3] J. Bernholc,[2,3,*] An-Ping Li[1,*]

[1]Center for Nanophase Materials Sciences, Oak Ridge National Laboratory, Oak Ridge, TN 37831, USA

[2]Department of Physics, North Carolina State University, Raleigh, NC 27695, USA

[3]Computational Sciences and Engineering Division, Oak Ridge National Laboratory, Oak Ridge, TN 37831, USA

*Email: apli@ornl.gov, bernholc@ncsu.edu

Precision control of interfacial structures and electronic properties is the key to the realization of functional heterostructures. Here, utilizing the scanning tunneling microscope (STM) both as a manipulation and characterization tool, we demonstrate the fabrication of a heterostructure in a single atomically precise graphene nanoribbon (GNR) and report its electronic properties. The heterostructure is made of a seven-carbon-wide armchair GNR and a lower band gap intermediate ribbon synthesized bottom-up from a molecular precursor on an Au substrate. The short GNR segments are directly written in the ribbon with an STM tip to form atomic precision intraribbon heterostructures. Based on STM studies combined with density functional theory calculations, we show that the heterostructure has a type-I band alignment, with manifestations of quantum confinement and orbital hybridization. Our finding demonstrates a feasible strategy to create a double barrier quantum dot structure with atomic precision for novel functionalities, such as negative differential resistance devices in GNR-based nanoelectronics.

## I. INTRODUCTION

Interfaces determine the key properties of electronic devices [1]. Precise control of interface structure presents a transformative opportunity for exploring novel device functionalities at the atomic scale. Graphene and related low-dimensional (LD) materials are a model system to explore the interfacial behavior and have been proposed as a promising material solution to computing beyond Moore's Law [2]. Recent advances in bottom-up synthesis of atomically precise graphene nanoribbons (GNRs) [3-23], as well as GNR-based heterostructures (HSs) formed by fusing two or more GNR segments together via on-surface reactions [24-29], provide unique opportunities to make genuine progress toward a bottom-up synthesis of functioning interfaces. Recently, it has been shown [21,30] that GNRs can be synthesized from polymer precursors by charge injections from a scanning tunneling microscope (STM) tip to trigger chemical reactions at selected molecular sites, thus enabling fabrication of designer heterojunctions with atomic scale precision. The applicability of this approach will hinge on demonstration of direct writing of atomically-precise interfaces with desired electronic properties. However, this promise is confounded by the fact that GNR edges, defects, confinement effects, and interactions with supporting substrate can all dramatically alter the electronic properties of these materials. It is thus pivotal to examine the electronic energy levels at the interface and explore routes to control their alignments for desirable interfacial functions.

In this paper, we demonstrate the direct writing of atomically precise HSs in single GNRs by the STM tip and study their interfacial electronic structures. The fabricated GNR-based HSs consist of seven-carbon-atom wide armchair GNRs (7-aGNRs) and an intermediate state in GNR synthesis, which is a partially-converted GNRs with one side of the polyanthrylene converted to the GNR structure while the other side remains in the polymeric structure. The STM is then used to characterize the atomic structure, energy level alignments, quantum confinement effects, and hybridization of electronic levels of the HSs. Our combined STM and density functional theory (DFT) study shows that the fabricated GNR/intermediate HS has a type-I band alignment. We then fabricate a double barrier quantum dot-like structure that is expected to display negative differential resistance (NDR). Our findings pave the way to controllable fabrication of GNR-based designer HSs with desired functionalities at the atomic scale.

## II. EXPERIMENTAL AND COMPUTATIONAL DETAILS

The 10,10'-dibromo-9,9'-bianthryl (DBBA) molecules were adopted as precursors to grow the 7-aGNRs on an Au(111) substrate, as reported previously [3,21,23,28]. Before deposition, the Au(111) single crystal was cleaned by repeated cycles of $Ar^+$ bombardment and annealing to 740 K. The DBBA molecules with a purity of 98.7% were degassed overnight at 450 K in a Knudsen cell. The molecule precursors were then evaporated from the cell at 485 K, while the Au substrate was held at 470 K. The molecules became dehalogenated upon adsorption. The sample was annealed at 470 K for 30 min to induce colligation/polymerization. In a subsequent step, annealing to 670 K will fully convert the polymer chains to the 7-aGNRs, but we chose a lower temperature of 600 K to induce partial conversion to the intermediate between the polymer and the 7-aGNR, as illustrated in Fig. 1(a).

STM characterization was performed with a home-made system at 105 K under ultrahigh vacuum conditions (base pressure better than $1 \times 10^{-10}$ Torr) with a well-cleaned commercial PtIr tip in a constant-current mode. The d$I$/d$V$ spectra were recorded using a lock-in amplifier with a sinusoidal modulation ($f = 731$ Hz, $V_{mod} = 20$ mV) with the feedback-loop gain off. The polarity of the applied voltage refers to the sample bias with respect to the tip.

DFT calculations were performed with the Quantum Espresso code [31], using ultrasoft pseudopotentials [32] with Perdew-Burke-Ernzerhof (PBE) exchange correlation potential [33]. The energy cutoff for the plane wave basis of Kohn-Sham wavefunctions was 30 Ry, and that for the charge density was 300 Ry. The atomic structures of the HSs were relaxed until forces on atoms reached a threshold of 0.002 Ry Å$^{-1}$ in vacuum. The calculations of the local density of states (LDOS) employed a Gaussian smearing of 0.01 eV.

### III. RESULTS AND DISCUSSION

By lowering the annealing temperature in the cyclodehydrogenation step, most of the polymers have been converted to the 7-aGNRs with some partially converted GNRs. The partially converted intermediate regions are marked with white boxes in Fig. 1(b). Note that more than one isolated segment of the intermediate can be formed in a single ribbon. Figure 1(c) shows the high-resolution STM image of an intermediate segment with one-side cyclodehydrogenation embedded in a 7-aGNR. On the cyclodehydrogenated side, a period of λ ~ 0.43 nm is observed, which is the same as the fully graphitized region, while the other side has a period of $\lambda_{int}$ ~ 2λ ~ 0.84 nm. The structural model is shown in the lower panel of Fig. 1(c), consistent with the one-side-cyclodehydrogenation model [30].

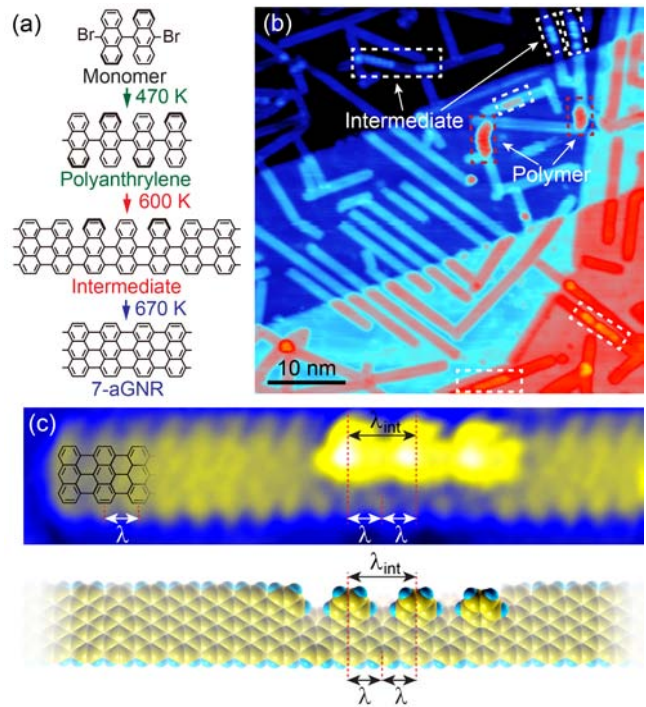

FIG. 1. (a) Temperature dependent growth process of bottom-up synthesis of the 7-aGNRs using DBBA molecules as precursors. After the formation of polyanthrylene, subsequent annealing at a temperature of 600 K will create an intermediate state with one-side cyclodehydrogenation while annealing to 670 K will fully convert the polymer chains to the 7-aGNRs. (b) Large-area STM image after partial graphitization at 600 K (Sample voltage $V_s$ = −2.0 V, tunneling current $I_t$ = 100 pA). Residual polymer and intermediate segments are marked with red and white dashed boxes, respectively. (c) Upper panel: High-resolution STM image showing the atomic structure of an intermediate ($V_s$ = −0.5 V, $I_t$ = 100 pA). Lower panel: Structural model of the intermediate.

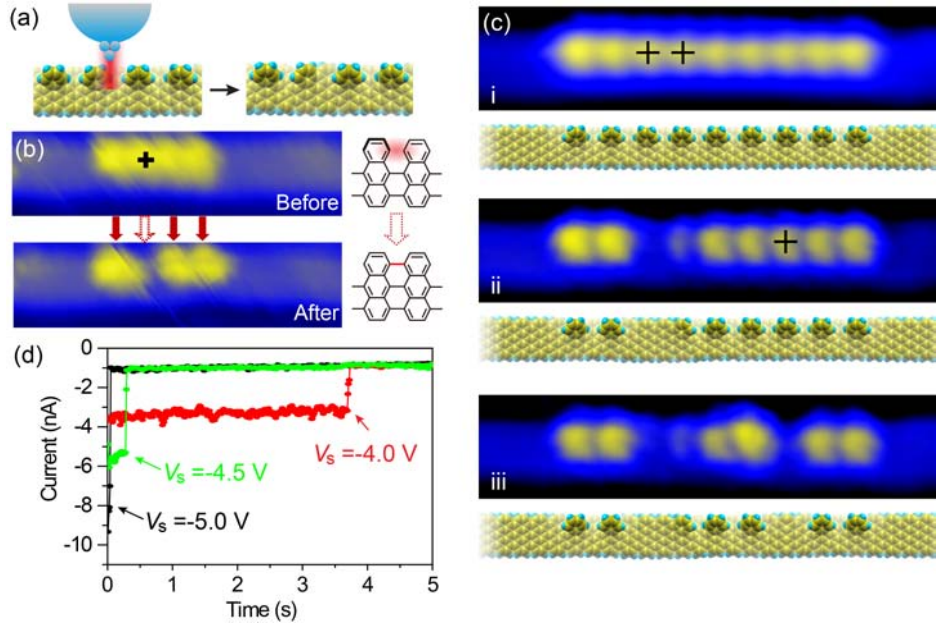

FIG. 2. (a) Schematic of the direct writing of HSs with an STM tip. (b) Left: STM images of a short intermediate segment with four units embedded in a 7-aGNR before and after a pulse treatment ($V_s = -2.0$ V, pulse time $t = 30$ ms) at the black-cross marked site ($V_s = -2.0$ V, $I_t = 100$ pA). Right: Schematic of the STM tip induced reaction marked by the dashed arrow. (c) Sequential direct writing of GNR segments in the intermediate to form HSs at designed molecular sites marked by crosses. Conversion of two neighboring intermediate units in (i) to a GNR segment in (ii), and sequentially of another unit of the intermediate in (ii) to a GNR segment in (iii) (setpoint: $V_s = -2.0$ V, $I_t = 20$ pA). The direct writing is enabled by STM pulses at $V_s = -5.0$ V, $I_t = 20$ pA, and time $t = 30$ ms. The corresponding structural models are shown in lower panels. (d) Representative $I$-$t$ curves at different biases with feedback loop off while the tip is held above the intermediates at setpoint of $V_s = -2.0$ V, $I_t = 100$ pA. The arrows mark the abrupt changes in the $I$-$t$ curves and the corresponding biases.

Although the GNR/intermediate HS can be formed by the two-step annealing process illustrated in Fig. 1(a), the obtained HSs are randomly distributed in GNRs.

To achieve better control of HS fabrication and to write an HS in a single GNR at a designed location, we now fabricate different HSs from the intermediate structures in a controllable manner, where an STM tip is used to "direct write" the designer interfaces. As we demonstrated in Ref. [21] and illustrated in Fig. 2(a), hole injections from an STM tip can be used to convert the polymer into GNR. The effect of tip treatment is shown in Fig. 2(b), where one unit cell of an intermediate segment is converted to GNR with a pulse of −2.0 V for 30 ms. Using this procedure, we can direct write multiple GNR/intermediate HSs at selected molecular sites in a single ribbon. Figure 2(c) shows the fabrication process of writing a 5-part GNR/intermediate HS in a long intermediate segment. By applying current pulses with a bias of −5 V for 30 ms at selected sites (marked with crosses), we can trigger the conversion of intermediate segments to GNR segments. Moving the tip along the intermediate and repeating this tip treatment, we can write GNR/intermediate HSs with more segments and designed segment lengths. Figure 2(d) shows three recorded *I-t* curves during pulse treatments at different biases. All curves show a current jump from a high plateau to a low plateau, indicating the conversion of the protruding intermediate to a planar GNR segment. The pulse time needed for this treatment is increased when the bias is decreased from −5 V to −4 V. Note that we write the HS from an intermediate segment instead of directly from a polyanthrylene chain. When writing GNRs directly from polymer chains with tip-induced conversion [21], we found that high biases easily broke the polymer backbone before inducing the conversion. In contrast, the preexisting C-C backbone structure on the graphitized side of the intermediate leads to better tolerance of the bias treatment, and thus allows better control in creating designer GNR/intermediate HSs.

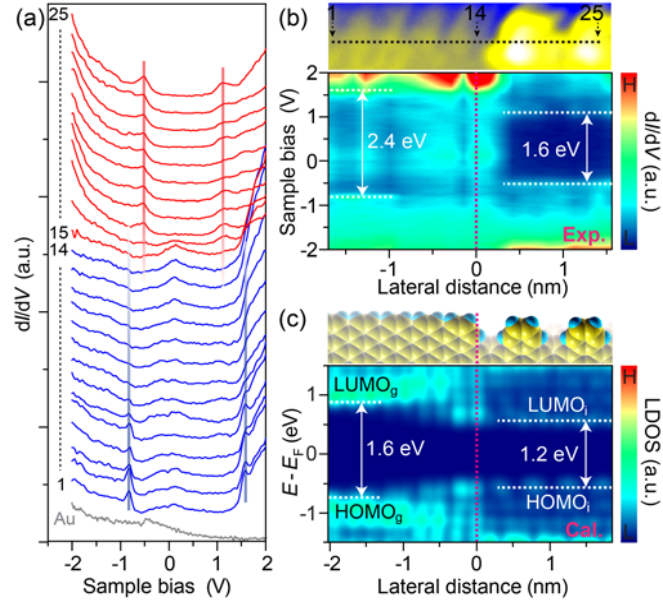

FIG. 3. (a) d$I$/d$V$ spectra acquired along the black dashed line across the interface between the 7-aGNR and an intermediate shown in the upper panel of (b). A gray curve is acquired on the bare Au surface as reference, showing a typical Au surface state at about −0.45 V. Blue and red curves are from the 7-aGNR and the intermediate segment, respectively. The HOMO and LUMO peaks for each segment are marked. (b) Upper panel: STM of a GNR/intermediate HS, showing the polymer side edge. Lower panel: Color-coded d$I$/d$V$ map plotted with the measured curves in (a). (c) Upper panel: Structural model showing the polymer side edge of the GNR/intermediate junction. Lower panel: Calculated LDOS map. The pink dashed line marks the junction location, and the white lines mark the HOMO and LUMO levels for corresponding segments.

We now characterize the electronic structures of the GNR/intermediate HS with scanning tunneling spectroscopy (STS). Figure 3(a) shows STS curves acquired across the GNR/intermediate junction. Away from the junction interface, the 7-aGNR exhibits the characteristic band gap of about 2.4 eV, with the highest occupied and lowest unoccupied molecular orbitals of the 7-aGNR ($HOMO_g$ and $LUMO_g$) at about −0.8 eV and 1.6 eV, respectively, consistent with previous reports [5,6,9,16,34]. The intermediate displays a smaller band gap of about 1.6 eV, with the $HOMO_i$ and $LUMO_i$ at about −0.5 eV and 1.1 eV, respectively. When approaching the interface, the $HOMO_g$ and $LUMO_g$ in the 7-aGNR are strongly suppressed, while the $HOMO_i$ and $LUMO_i$ in the intermediate are barely affected. The electronic structure evolution across the 7-aGNR/intermediate interface is visualized in a 2D color-coded LDOS map shown in Fig. 3(b). For comparison, the calculated LDOS map is shown in Fig. 3(c), where the $HOMO_i$ and $LUMO_i$ of the intermediate are found to extend to the 7-aGNR segment by about 1 nm. Clearly, both the STS measurement and the LDOS calculations show a smaller band gap for the intermediate than the 7-aGNR with a type-I band alignment. STS measurements give 1.6 eV versus 2.4 eV; the theoretical values being 1.2 eV versus 1.6 eV. Note that the simulated band gaps are smaller than the experimental ones due to underestimation of band gaps in DFT calculations [35].

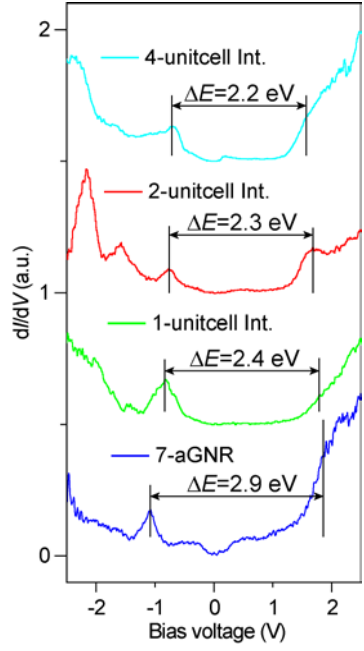

FIG. 4. d$I$/d$V$ curves taken on the intermediate (Int.) segments with different lengths as shown in Fig. 2(b), in comparison with a curve from the 7-aGNR. The band gaps are labeled to show a confinement effect.

The band gap size of the intermediate varies with length due to quantum confinement effect. Figure 4 shows STS curves from three intermediate segments with lengths of 1, 2 and 4 unit cells embedded in the same GNR, showing respective band gaps of 2.4, 2.3 and 2.2 eV. While the band gaps of GNRs and the intermediates depend on variations of the confinement and substrate screening [16,21,34,36], the intermediates always have a smaller band gap than the corresponding 7-aGNR segment within the same ribbon, giving consistently type-I band alignment. As the 7-aGNR has a larger band gap, this results in the conduction band of the 7-aGNR being higher than that of the intermediate and the valence band being lower. One can thus use the 7-aGNR as a barrier and the intermediate segment as a quantum dot in a double barrier NDR device [37-39].

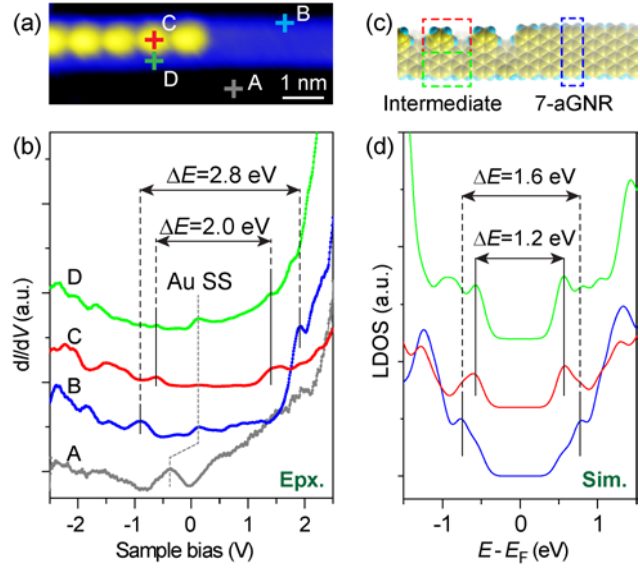

FIG. 5. (a) STM image showing 7-aGNR/intermediate HS ($V_s = -2.0$ V, $I_t = 40$ pA). (b) d$I$/d$V$ curves acquired at cross-marked sites in (a): Curve A (gray) is measured on the bare Au surface, Curve B (blue) on the fully converted 7-aGNR edge, Curve C (red) on the polymer side of the intermediate, and Curve D (green) on the graphitized side of the intermediate. (c) Structural model of the intermediate. (d) Calculated LDOSs correspondingly obtained from the dashed boxes in (c).

The polymeric features within the intermediate segment appear like defects in the single 7-aGNRs [40], but they do not possess localized defect states. Figure 5(a) shows a high resolution STM image of a GNR/intermediate HS, with the corresponding d$I$/d$V$ curves shown in Fig. 5(b) acquired at marked locations shown in Fig. 5(a). The same band gap is found for the polymer (Curve C) and the graphitized sides (Curve D) of the intermediate. The d$I$/d$V$ curves from the pristine 7-aGNR segment (Curve B) and the partially graphitized side of the intermediate (Curve D) both show contributions from the Au surface states (Curve A), but the band gap of the Curve C is clearer due to suppression of the Au surface states by the out-of-plane feature of the polymer side. The uniform band gap in the intermediate is well reproduced by DFT calculations shown in Fig. 5(d). It is a consequence of strong hybridization between the tilted benzyne rings at the polymer side and the narrow GNR structure at the graphitized side. Interestingly, although the intermediate contains a narrower GNR-like structure [marked with the green dashed box in Fig. 5(c)] than the 7-aGNR, it has a band gap smaller than the 7-aGNR. In a previous report [30], the graphitized side of the intermediate was named $5^+$-aGNR; however, although the intermediate structure contains the backbone of a 5-aGNR, its electronic properties are significantly different from the nearly metallic behavior of the intrinsic 5-aGNR [7].

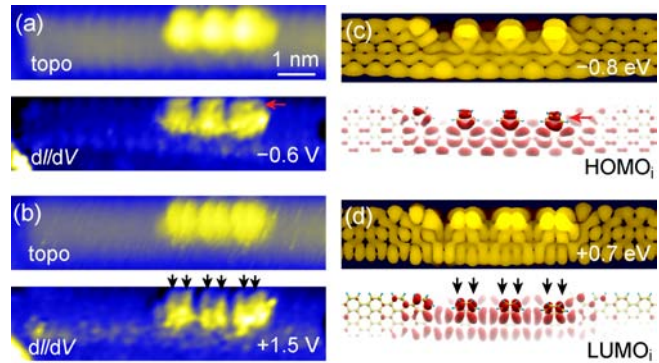

FIG. 6. (a) and (b): STM topographic images (upper panels) and simultaneously acquired d$I$/d$V$ maps (lower panels) at −0.6 V and +1.5 V, respectively ($I_t$ = 100 pA). (c) and (d): Simulated STM images (upper panels) at marked energies and charge density distributions (lower panels) at the energies of HOMO$_i$ and LUMO$_i$ of the intermediate, respectively. Sub-hexagon resolved structures are marked with red and black arrows. HOMO$_i$ and LUMO$_i$ are both extended in the intermediate in (c) and (d).

The strong orbital hybridization in the intermediate can be more clearly seen in the spatial distribution maps of the HOMO and LUMO states. Figures 6(a) and 6(b) show the simultaneously acquired topographic images and d$I$/d$V$ maps at −0.6 and +1.5 V, respectively, correspondingly to the HOMO$_i$ and LUMO$_i$. The HOMO$_i$ [Fig. 6(a), lower panel] displays a bright round feature with a short (one anthrylene width) stripe along the edge (marked with a red arrow), which turns into two perpendicular stripes (marked with black arrows) in the LUMO$_i$ [Fig. 6(b), lower panel]. These features are well captured by the simulated HOMO$_i$ and LUMO$_i$, respectively, as shown in Figs. 6(c) and 6(d). Due to the out-of-plane structures, the STM images of HOMO$_i$ and LUMO$_i$ both display higher contrast at the polymer side in the intermediate. However, the corresponding charge densities are distributed much more uniformly in the intermediate for both energies [Figs. 6(c) and 6(d)], which agrees with the experimental data in Fig. 5. Therefore, the polymeric features at one side of the intermediate cannot be treated as "isolated" structural defects; they are electronically hybridized with the GNR backbone and the intermediate needs to be considered as a single segment forming an intraribbon junction with the 7-aGNR. When built into a device, the intermediate will not act as a localized scattering center. Instead, it will serve as a functional component, e.g., a quantum dot confined by GNR barriers in a NDR device [41].

## IV. SUMMARY AND CONCLUSIONS

By using the STM tip as a manipulation tool, we show that atomically precise heterostructures can be written directly in single GNRs. The heterostructures consist of seven-carbon-atom wide armchair GNRs and an intermediate structure that has only one side of the polyanthrylene converted to the GNR structure while the other side remains in the polymeric structure. Multiple junctions can be written at selected molecular sites to build a functional device structure. The heterostructures and their electronic properties are characterized using an STM and DFT calculations. It is found that the junction has a type-I band alignment and that the intermediate segment embedded in a single GNR behaves like a quantum dot confined in a double barrier structure. The fabricated GNR/intermediate HS is expected to display negative differential resistance.

## ACKNOWLEDGMENTS

A portion of this research was conducted at the Center for Nanophase Materials Sciences (CNMS), which is a DOE Office of Science User Facility. The research was funded by grants ONR N00014-16-1-3213 and N00014-16-1-3153, and DOE DE-FG02-98ER45685. The development of the RMG code was funded by NSF grant OAC-1740309. Supercomputer time was provided by NSF grant ACI-1615114 at the National Center for Supercomputing Applications (NSF OCI-0725070 and ACI-1238993).

[#]C.M. and Z.X. contributed equally to this work.